\documentclass[%
 aip,
 superscriptaddress,
 apl,%
 amsmath,amssymb,
 reprint,%
]{revtex4-2}
\usepackage{siunitx}
\usepackage{graphicx}% Include figure files
\usepackage{dcolumn}% Align table columns on decimal point
\usepackage{bm}% bold math
\begin{document}		
\title{Enhanced photovoltaic effect in graphene-silicon Schottky junction under mechanical manipulation}
\date{\today}

\author{Dong Pu}
\thanks{These two authors contributed equally }
\author{Muhammad Abid Anwar}
\thanks{These two authors contributed equally }
\author{Jiachao Zhou}
\affiliation{School of Micro-Nano Electronics, Zhejiang University, Hangzhou, China}
\affiliation{ZJU Global Scientific and Technological Innovation Center, Hangzhou, China}
\author{Renwei Mao}
\author{Xin Pan}
\affiliation{ZJU-UIUC Institute (ZJUI), Zhejiang University, Haining, China}
\author{Jian Chai}
\author{Feng Tian}
\author{Hua Wang}\thanks{Corresponding author}\email{daodaohw@zju.edu.cn}
\affiliation{School of Micro-Nano Electronics, Zhejiang University, Hangzhou, China}
\affiliation{ZJU Global Scientific and Technological Innovation Center, Hangzhou, China}
\author{Huan Hu}\thanks{Corresponding author}\email{huanhu@intl.zju.edu.cn}
\affiliation{School of Micro-Nano Electronics, Zhejiang University, Hangzhou, China}
\affiliation{ZJU-UIUC Institute (ZJUI), Zhejiang University, Haining, China}

\author{Yang Xu}\thanks{Corresponding author}\email{yangxu-isee@zju.edu.cn}
\affiliation{School of Micro-Nano Electronics, Zhejiang University, Hangzhou, China}
\affiliation{ZJU Global Scientific and Technological Innovation Center, Hangzhou, China}
\affiliation{ZJU-UIUC Institute (ZJUI), Zhejiang University, Haining, China}

\begin{abstract}
Graphene-silicon Schottky junction (GSJ) which has the potential for large-scale manufacturing and integration can bring new opportunities to Schottky solar cells for photovoltaic (PV) power conversion. 
However, the essential power conversion limitation for these devices lies in the small open-circuit voltage ($V_{oc}$), which depends on the Schottky barrier height (SBH).
In this study, we introduce an electromechanical method based on the flexoelectric effect to enhance the PV efficiency in GSJ.
By atomic force microscope (AFM) tip-based indentation and in situ current measurement, the current-voltage (I-V) responses under flexoelectric strain gradient are obtained. 
The $V_{oc}$ is observed to increase for up to 20$\%$, leading to an evident improvement of the power conversion efficiency.
Our studies suggest that strain gradient may offer unprecedented opportunities for the development of GSJ based flexo-photovoltaic applications.

\end{abstract}

\maketitle

Due to its special two-dimensional structural properties and excellent performance \cite{Geim2007}, graphene has the potential to be integrated into existing semiconductor technologies and used in next-generation electronics. 
Recent research has shown that the formation of junctions between graphene and three-dimensional or two-dimensional semiconductors \cite{xu2016contacts} can produce the rectification effect of a typical Schottky junction. 
Its tunable Schottky barrier makes graphene junctions an excellent platform for studying the transport properties of interfaces and has led to applications in scenarios such as photodetection \cite{Wang2013a}, light-speed communication \cite{Li2015}, and chemical and biological detection \cite{He2012}.
Solar energy collection and conversion have attracted much attention in recent years. 
The Schottky junction devices can naturally work as solar cell or photovoltaic cell \cite{Li2010c}, as the built-in electrical field provides the voltage potential difference that drives the current \cite{DiBartolomeo2016}. 
Conventional metal/semiconductor Schottky devices suffer from the contact instability, high cost and high-temperature fabrication process. 
Graphene, which has unique optical properties and excellent mechanical properties, offers Schottky solar cells with low sheet resistance, high optical transparency, large area growth, and low-cost transferring.
Over the last decade, many studies have been focused on Gr/Si Schottky junction \cite{doukas2022thermionic,ono2022thermal,wu2022vertical,geng2022directly}  (GSJ) for solar cell applications. 
An overall power conversion efficiency (PCE) of 1-1.7$\%$ is achieved with open circuit voltage ($V_{oc}$) and short circuit current ($J_{SC}$) linearly depending on the intensity of incident light\cite{Li2010c}. 
By chemical doping, PCE can be further increased for $8.5\%$ \cite{Chen2011}.
With systematical optimization including the number of graphene layers, PCE can surpass $3\%$ \cite{An2013}.
Furthermore, substrate metasurface \cite{zhang2013high,wu2013graphene}, such as Si nanowire, Si nanohole array, and additional antireflection layer \cite{shi2013colloidal} are proposed for PCE enhancement. 
\begin{figure*}[htb]
  \includegraphics[width=0.95\linewidth]{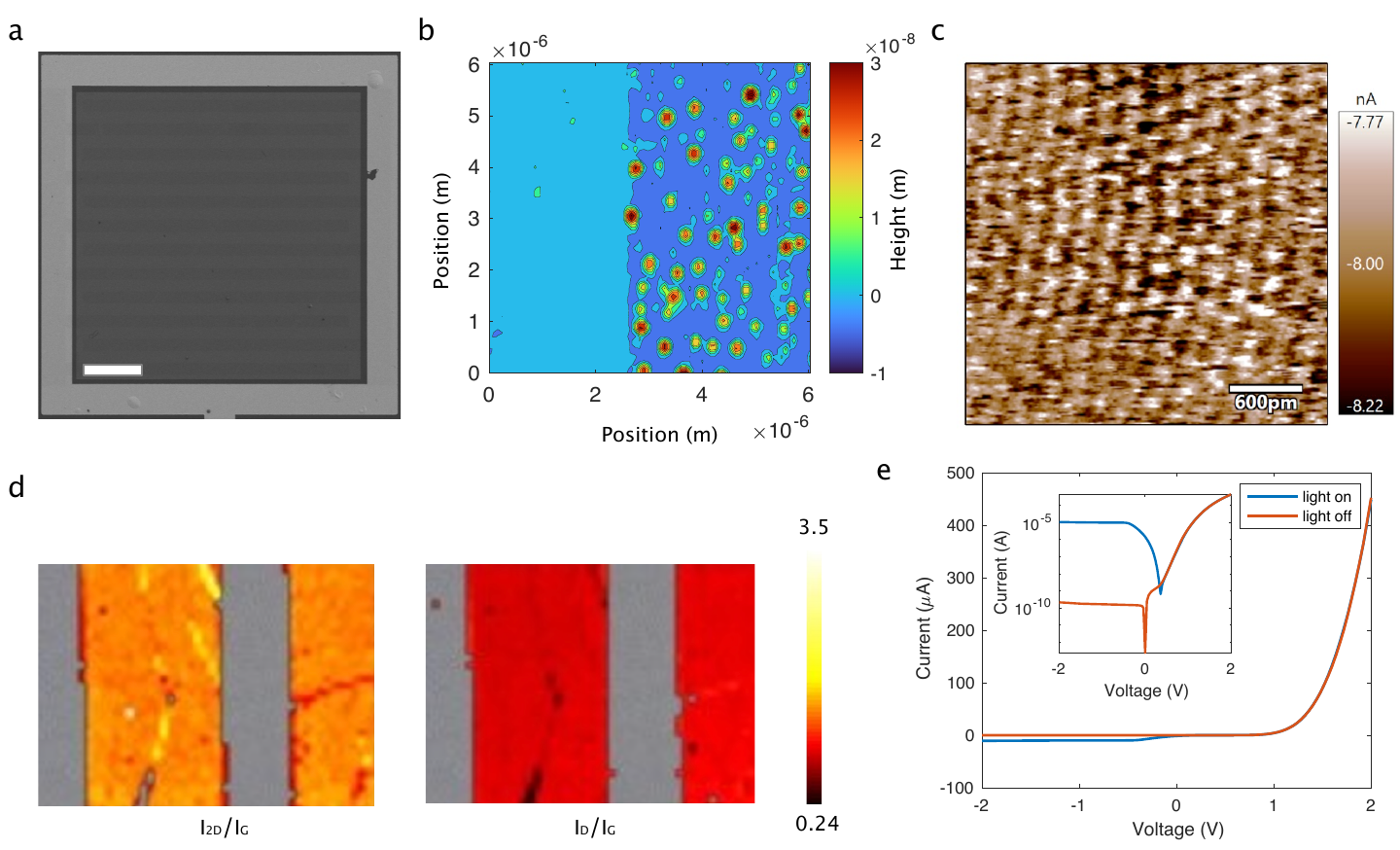}
  \caption{Experimental characterization of the GSJ device. 
  a. The Scanning electron micrograph (SEM) of the device. The white bar is \SI{100}{\mu m}.
  b. The surface topography of the device measured by AFM.
  c. Current distribution of Gr/Si junction measured by AFM with bias voltage $V_{bias}$\SI{-1}{V} and applied force \SI{0.26}{\mu N}.  The white scale bar represents \SI{600}{pm}.
  d. The Raman mapping of the device.
  e. The current-voltage response curve. The inset picture shows the logarithmic current as a function of the bias voltage. 
  }
    \centering
  \label{fig1}
\end{figure*}

However, one primary limitation for the GSJ Schottky solar cell is the low open circuit voltage $V_{oc}$, which relates to the small Schottky barrier height (SBH). 
Here, we introduce an electromechanical coupling, called flexoelectricity \cite{Wang2019h}, to increase the SBH, thus enhancing the performance of GSJ solar cell. 
The flexoelectric effect describes the linear coupling between electric polarization and strain gradient in solid state materials \cite{kogan1964piezoelectric,Tagantsev1986}. 
It suggests that the polarization can originate from the strain gradient even in the centrosymmetric systems.
Compared with the piezoelectric effect that has been studied extensively, the response of the flexoelectric effect is very weak, and remains underexplored \cite{Wang2019h}. 
The electric polarization induced by a strain gradient is typical of the order of \SI{e-9}{C/m^2} \cite{Wang2019h,Wang2020o} at the macro scale.
As the geometric size scales down, the strain gradient is inversely proportional to the spatial scale \cite{Park2021a}.
Thus micro-nano structure is able to achieve a large strain gradient, the flexoelectric effect induced by a strain gradient, and dominates over the piezoelectric effect at nanoscale\cite{Naumov2009,Springolo2020}.
In recent years, By using an atomic force microscope (AFM) to introduce large strain gradients, a number of experimental studies concentrate on the mechanism of flexoelectricity and its applications have emerged \cite{Park2021,Jiang2022,Wang2021e}. 
Conductive AFM (CAFM) probe coated with metal can introduce a strain gradient and simultaneously monitor the current flow through the junctions. 
The strain gradient breaks the inversion symmetry of centrosymmetric materials and induces electric field polarization, named flexoelectric effect \cite{Wang2020o}.
In contrast, a uniform strain cannot induce dipoles in graphene \cite{wang2015observation}, and it is difficult to change the Schottky barrier.
The barrier height of the Schottky junction interface between the probe and silicon can be tuned by the flexoelectric effect \cite{Wang2020o,Sun2021}.
The flexo-photovoltaic effect (FPV) was found in perovskites \cite{Yang2018,Shu2020}
and two-dimensional material system \cite{Zhang2019d,Jiang2021}, and significantly improves the solar cell performance. 
These motivate us to systematically investigate the flexo-photovoltaic in GSJ.

In this study, we introduce the flexo-photovoltaic effect in the Gr/Si Schottky junction.
We find the GSJ performance as a solar cell can be largely enhanced through the flexoelectric effect by using AFM tip pressing.
By in situ adding mechanical force on the GSJ, the current flows through the junction to the tip can be detected, then being read out based on the CAFM module using AFM.
The current-voltage curves under different applied forces are analyzed, and we obtained the corresponding SBH variation as a function of applied force.
Under illumination, the GSJ device shows PV effect. 
$V_{oc}$ can be improved by this electromechanical effect.
We finally demonstrate the enhanced PV through the flexoelectric effect in GSJ. 

The graphene/silicon junction Schottky devices are prepared with 
a single layer of CVD graphene being wet-transferred to the lightly-doped n-type silicon layer forming the two dimensional and three dimensional (2d-3d) Schottky contact.
An n-doped Si/SiO$_2$ \SI{500}{\mu m}$/$\SI{100}{\mu m} substrate with a resistivity of 1-\SI{10}{\Omega \cdot cm} corresponding to a doping concentration of \SI{4.5e14}{cm^{-3}}  to \SI{4.94e15}{cm^{-3}} is used.
%An n-doped Si/SiO$_2$ (\SI{500}{\mu m}$/$ \SI{100}{\mu m} substrate with a resistivity of \SI{1−10}{\Omega \cdot cm} corresponding to a doping concentration of \SI{4.5e14}{cm^(-3)}  to \SI{4.94e15}{cm^(-3)} is used.

The bottom of the silicon substrate is mechanically scraped to remove the thin oxide layer and coated with GaIn and copper to form an ohmic contact \cite{Peng2022a}.
The scanning electron micrograph (SEM) of the device is shown in Fig.\ref{fig1}a.
The graphene covers the silicon window (dark area) forming the Schottky contact, connecting with the Au electrode (surrounding gray part), and forming an ohmic contact with Au electrodes.
The graphene is etched into ribbons.
The surface morphology of the Gr ribbon edge tested by AFM is shown in Fig.\ref{fig1}b.
The left-hand side flatten area is silicon covered by graphene, while the other is bare silicon after the graphene is etched. 
With the tip contacting the GSJ area, the current can be read out with the CAFM module. 
Figure.\ref{fig1}c shows the high spatial resolution current distribution of the junction with $V_{bias}=$ \SI{-1}{V}. 
Figure.\ref{fig1}d shows the corresponding Raman mapping on the graphene ribbon device.
\begin{figure*}[bt] 
  \centering
  \includegraphics[width=0.75\linewidth]{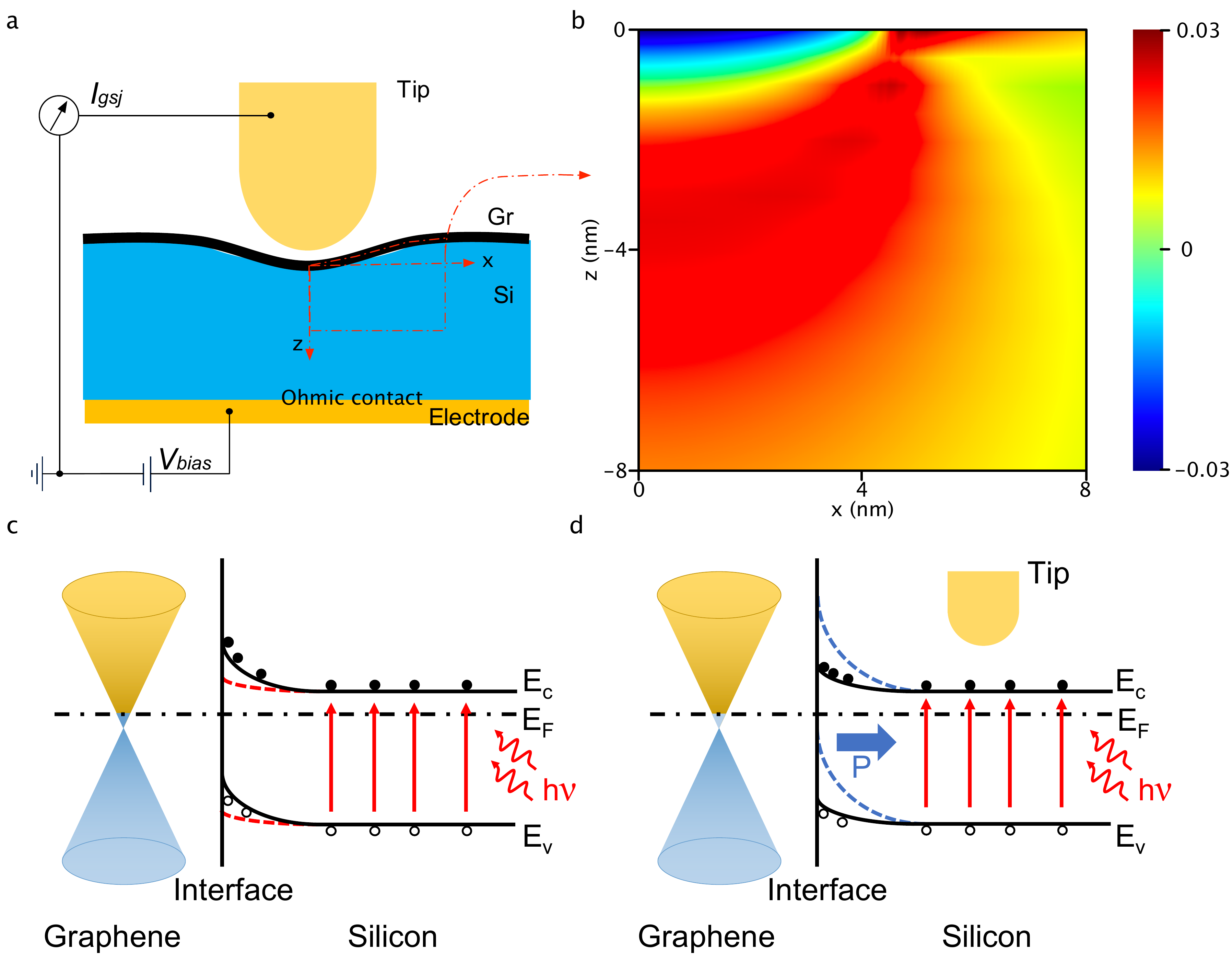}
  \caption{a. Schematic of the CAFM experimental setup.
  b. Finite element analysis (FEA) of strain distribution in the substrate under pressing.
  c. Energy bandgap structure of the GSJ with laser on.
  d. The bandgap structure bending by tip-pressing induced flexoelectricity modulation.
  }  
    \label{f2}
\end{figure*}
The intensity of the 2D peak versus the G peak and the intensity of the D peak versus the G peak are shown to illustrate the clear graphene ribbon areas. 
According to the contour scale bar, the 2D peak has a larger peak, while the D peak has a smaller peak, indicating the high quality of the graphene in the sample.
With the probe connecting the bottom copper and the top-surface electrode separately, we obtain the basic current-voltage (I-V) response of the device using a semiconductor analyzer (Agilent B1500), shown in Fig.\ref{fig1}e.
The red line shows I-V response with the light turned off while the blue line represents the light-on case. 
In the light-off case, the I-V curves show the typical Schottky diode properties of GSJ.
The center wavelength is \SI{532}{nm} and the approximate power density used is \SI{50}{\mu W}.
The reverse current is on the order of \SI{e-10}{A}, showing a good diode characteristic. 
When the light turns on, the reverse photocurrent exceeds \SI{10}{\mu A}.
The device performs a typical photoresponse of GSJ, and the inset picture shows the half-logarithmic curve of the I-V results.
In the reverse bias region, the photocurrent reaches nearly 5 orders of magnitude under the effect of the photoresponse. 
In the following experiments, we use AFM to do further tests to study the electromechanical effects on the responses by in situ exerting mechanical stress.
Note that due to the current-limiting protection of the instruments (\SI{20}{nA}), we focus on the results in the range of \SI{+-20}{nA} in the following tests.

Considering the graphene massless carrier property and the inhomogeneity of the Gr/Si contact, the modified equation for GSJ based on thermal emission theory can be expressed as the following formula\cite{DiBartolomeo2016,Liang2017},
\begin{equation}
   J=A^{*}T^3\exp \left(-\frac{\bar{\phi}_B-\frac{\delta_P^2}{2k_BT}}{k_BT} \right)[\exp \left(\frac{qV}{\eta k_BT}-1 \right) ]
   \label{jiv}
\end{equation}
where $A^{*}=$ \SI{0.01158}{A/cm^2/K^3} and $\delta_P =$ \SI{135}{meV}, and $\eta$, $k_B$, $T$ and $q$ are the ideal factor, the Boltzmann constant, temperature and elementary charge, respectively.
The effective working area $A$ of GSJ for our device here is \SI{1.25e-3}{cm^2}.
Fitting Fig.\ref{fig1}e with Eq.\ref{jiv}, we then obtain the Schottky barrier height (SBH) $\bar{\phi}_B=$ \SI{0.6}{eV} before applying mechanical stress. 
To find the effect of electrical polarization on the Schottky junction induced by strain gradient, we use the Conductive-AFM tip to exert stress on the graphene surface and in situ read-out the current ($I_{gsj}$) by sweeping the bias voltage $V_{bias}$, the schematic of the experimental setup is shown in Fig.\ref{f2}a.
\begin{figure*}[htbp]
  \centering
  \includegraphics[width=0.75\linewidth]{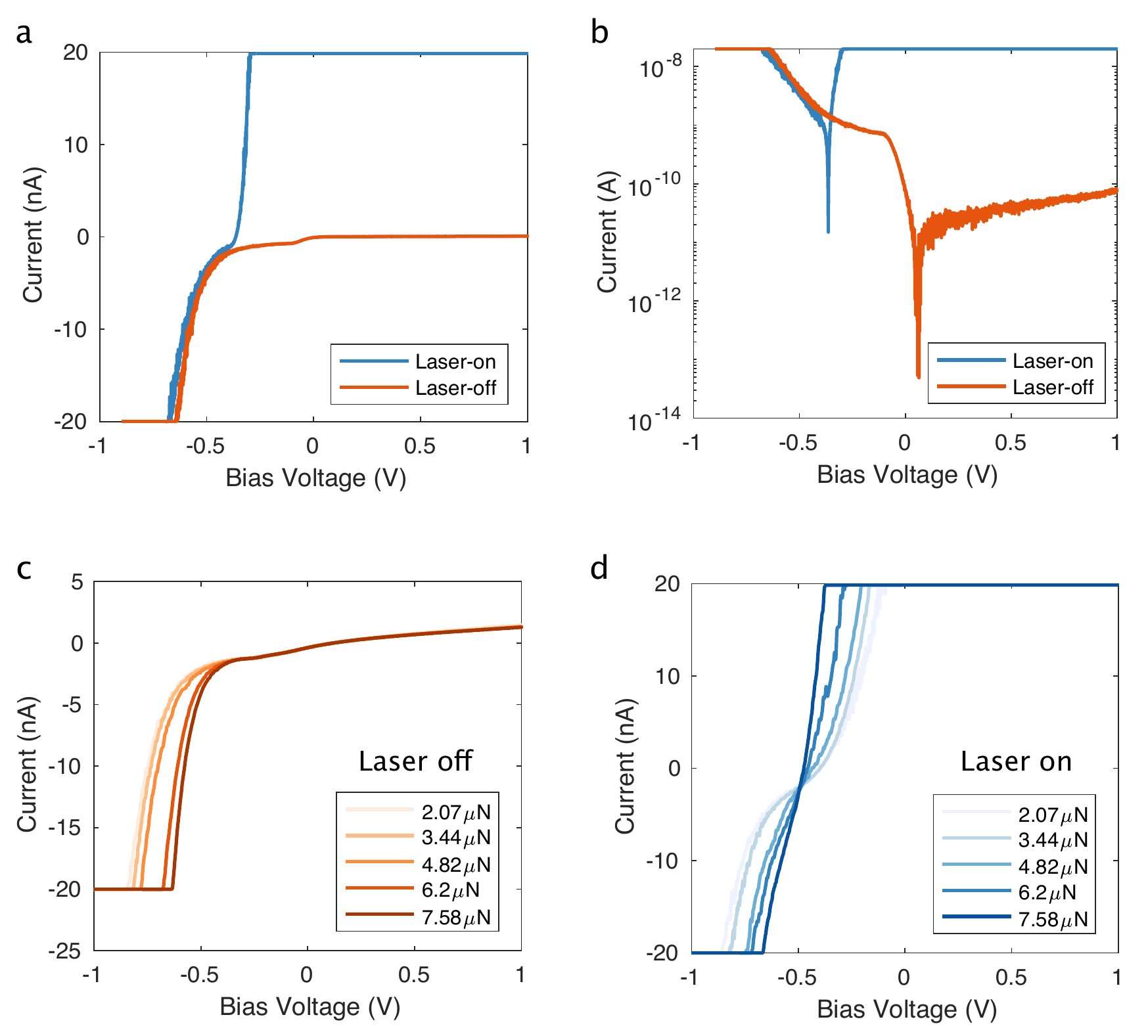}
  \caption{
  GSJ I-V responses under tip-pressing obtained by AFM.
  Photo-response effects on the GSJ performance with fixed applied force (a. Linear ordinate coordinates; b. Logarithmic ordinate coordinates).
  c. GSJ I-V responses of increased applied forces with laser turned off.
  d. GSJ I-V responses of increased applied forces with laser turned on.
  }  
  \label{f3}
\end{figure*}

The tip coated with conductive diamond (Adama-AD40) is directly pressed onto the CVD graphene layer.
Note that, in these AFM based experiments, $V_{bias}$ is applied to the bottom ohmic electrode and the current reversely flows through the Gr/Si Schottky junction to the probe for current detection. 
Thus the forward bias response is under the condition of negative bias voltage ($V_{bias}<0$).
Directly experimental calibration of the tip-induced strain distribution is still  challenging \cite{Park2021}. 
In the following, we estimate the strain distribution based on finite element simulation via COMSOL.
As shown in Fig.\ref{fig1}, the graphene is etched into ribbons.
Due to the atomic thickness of graphene and the free-standing edge of the ribbon, the graphene layer has negligible effects on the out-of-plane deformation of the junction. 
The finite element analysis on the strain distribution under pressing thus only takes the tip and the silicon substrate into consideration for simplification. 
We set the tip radius to \SI{10}{nm} according to the SEM imaging measurement.
The bottom of the substrate is set to be fixed according to the actual situation of the experiment.
The tip is set using diamond material with density \SI{3515}{kg/m^3}, Young's modulus \SI{105e10}{Pa} and Poisson's ratio 0.1 while the silicon substrate with density \SI{2329}{kg/m^3}, Young's modulus \SI{170e9}{Pa} and Poisson's ratio 0.28.
The half-cross-section of the substrate is shown in Fig.\ref{f2}b, as the contact area has rotation symmetry of z-axis.  
While at an applied force on the order of $\sim$\SI{5}{\mu N}, the corresponding strain is about \SI{+-0.3}{}.
For the vertical direction (z-axis), the strain gradient is estimated to be on the order of \SI{e7}{m^{-1}}.

For a typical graphene low-doped n-type silicon junction, the electronic band structure and energy-band diagram are shown in Fig.\ref{f2}c.
While under illumination, part of the photo-generated carriers moves to the graphene thus the barrier height slightly changes.
Note that in our following experiments, the laser central wavelength from AFM is \SI{840}{nm}.
In Fig.\ref{f2}d, we show the further changes (dashed blue lines) of the barrier height.
While under tip pressing, the strain gradient in the substrate silicon breaks the inverse symmetry of the centrosymmetric material, which results in a flexoelectric effect \cite{Yang2018,Wang2020o} (the blue arrow represents the polarization). 
The corresponding built-in electric field thus increases the SBH. 
This leads to the flexoelectricity induced Schottky barrier height tuning of the device, which will consequently affect the photovoltaic effect in this junction.
\begin{figure*}[tb]
  \centering
  \includegraphics[width=0.8\linewidth]{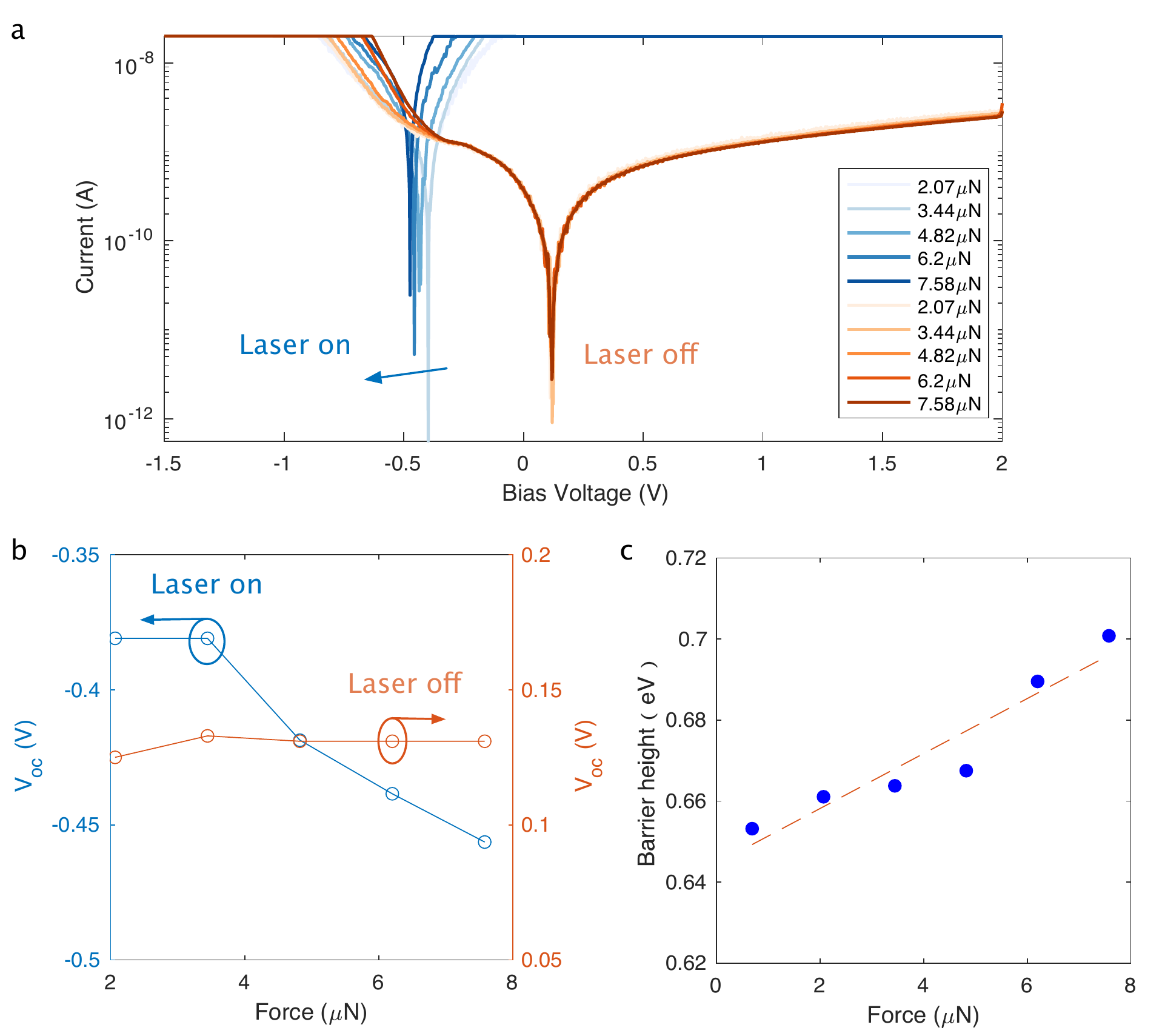}
  \caption{
  a. The GSJ current as a function of bias voltage under different applied forces. The orange curves show the laser-off cases while the blue curves show the laser-on cases. 
  b. The open-circuit voltage as a function of applied forces. The blue points represent the laser-on cases while the orange points represent the  laser-off cases.
  c. The extracted Schottky barrier height as a function of applied forces. 
  }  
  \label{f4new}
\end{figure*}

Figure \ref{f3} shows the GSJ I-V responses obtained directly by AFM while applying forces.   
While the laser is turned off, the GSJ shows a typical Schottky diode feature (the orange line in Fig.\ref{f3}a).
As aforementioned, the forward bias range is when $V_{bias}<0$, while the reverse bias area $V_{bias}>0$.
While $V_{bias}>$\SI{-0.3}{V}, the response current reaches the upper testing limit of our AFM instrument.
The open-circuit voltage ($V_{oc}$) increases to $\sim$\SI{0.4}{V}, shown clearly in Fig.\ref{f3}b.
The short-circuit current ($J_{sc}$) can not be directly obtained, because of the instrument current limits.
Nevertheless, the photoresponse (in Fig.\ref{f3}a and Fig.\ref{f3}b) implies that  this GSJ device has an evident photovoltaic effect. 
To further investigate the mechanical manipulation of the PV effect, we applied forces by AFM nanoindentation on the GSJ. 
We first turn the laser off to study the force induced effects on the GSJ. 
The response current (absolute value) in the forward bias area can be enlarged by increasing the applied force (Fig.\ref{f3}c) from \SI{2.07}{\mu N} to \SI{7.58}{\mu N}, while the response current in reverse bias area maintains being cut off.
Similarly, for each applied force, we then collect the current data by sweeping the bias voltage $V_{bias}$ when turning the laser on, shown in Fig.\ref{f3}d.
In the forward bias area, the GSJ current ($I_{GSJ}$) maintained being enlarged (absolute value) when the applied force is increased in contrast to the Fig.\ref{f3}c.
Due to the laser-induced photoresponse, the current dramatically changed. 
There exists an evident photovoltaic area. 
By directly contrasting the I-V responses under different applied forces, the half-logarithmic graph of the absolute GSJ current as a function of the bias voltage is shown in Fig.\ref{f4new}a.
For the laser-off cases (orange lines), the open-circuit voltage $V_{oc}$ remains stable  under different applied forces. 
Remarkably, for the laser-on cases (blue lines), due to the existing photovoltaic effects, $V_{oc}$ varies from near zero to around \SI{-0.5}{V} in contrast to the laser-off cases.
The absolute value of $V_{oc}$ is increased from \SI{0.38}{V} to \SI{0.46}{V} for more than 20$\%$, as shown in Fig.\ref{f4new}b (the blue circles). 
Fitting the laser-off cases with Eq.\ref{jiv}, the Schottky barrier height as a function of applied forces is obtained, as shown in Fig.\ref{f4new}c.  
The strain gradient in the substrate originates from the contacts between the tip and the substrate surface \cite{Park2021}, which breaks the inversion symmetry in the silicon \cite{Wang2020o,Sun2021} and causes an extra flexoelectricity induced built-in potential in addition to the native depletion layer in the GSJ.
The SBH can be additionally increased by the forces.

For this GSJ, the open-circuit voltage can be obtained for $I=0$ while short-circuiting current for $V=0$.
When $V=0$, the simple expression can be obtained \cite{DiBartolomeo2016} $I_{sc}\approx -I_{ph}$, where $I_{ph}$ is the photogenerated current. 
The $V_{oc}$ can be expressed as,
\begin{equation}
  V_{oc}\approx\frac{\eta k_BT}{q}\ln(\frac{I_{ph}}{I_0}) \approx\frac{\eta }{q}\phi_B+ Const.
\end{equation}
The $V_{oc}$ has approximately a linear relation with the Schottky barrier height, and agrees well with experiments. 

Considering the photoresponse of Gr/Si Schottky junction, we need the short-circuit current $I_{sc}$ and fill factor $FF$ for estimation. 
However, because of the limitation of our instrument, the short-circuit current can not be precisely measured experimentally. 
Here, our laser power has not been changed as an AFM detection light source for the laser-on cases, i.e. photogenerated current $I_{ph}$ is fixed.
We then assume that the short-circuit current for different applied forces are equal, i.e. for each case, when $V_{bias}=0$, the GSJ reaches the reverse saturated state.
In addition, when $V_{bias}> V_{oc}$, the current rises sharply to the upper limit in a series of parabolic shapes.
Considering the actual $I_{sc}\gg $\SI{20}{nA}, the fill factors $FF$ are assumed to be approximately equal for the cases under different forces. 
Consequently, the power conversion efficiency (PCE) can be expressed as $\displaystyle PCE=\frac{I_{sc}V_{oc}FF}{P_{in}}$, which can be improved for about at least 20$\%$ attributed to the enhancement of $V_{oc}$ under strain gradient. 

In this article, 
we investigate the flexoelectricity-enhanced photovoltaic effect in GSJ.
Using the AFM, we apply different forces to generate the strain gradient in the junction and in situ obtain the response current of GSJ under these conditions. 
By experimental validation, we find that the open-circuit voltage ($V_{oc}$) can be  enhanced from \SI{0.38}{V} to \SI{0.46}{V} by the applied forces when the laser is turned on and the Schottky barrier height can be increased from \SI{0.65}{eV} to \SI{0.7}{eV}, respectively.
As a consequence, the power conversion efficiency can be enhanced for more than 20$\%$ with the assumption that the short-circuit current $I_{sc}$ is identical and similar fill factor $FF$.
Our work shed light on the GSJ devices for PV effect enhancement with mechanical manipulation approach and pave the way for the PV enhancements 2D heterostructure based on the similar mechanism. At present, our research is still in the early stage of exploration, and combining this technology with integration is still a challenge for future applications.

\begin{acknowledgments}
This work was supported in part by the National Natural Science Foundation of China under Grants 92164106 and 61874094, China Postdoctoral Science Foundation (2021M692789), and in part by the Fundamental Research Funds for the Central Universities under Grants K20200060 and 2021FZZX001-17. We also thank Prof. Chunli Zhang for his valuable discussions and comments.
\end{acknowledgments}

\bibliography{FPV_in_GSJ_APL}

\begin{thebibliography}{34}
\providecommand{\natexlab}[1]{#1}
\providecommand{\url}[1]{\texttt{#1}}
\expandafter\ifx\csname urlstyle\endcsname\relax
  \providecommand{\doi}[1]{doi: #1}\else
  \providecommand{\doi}{doi: \begingroup \urlstyle{rm}\Url}\fi

\bibitem[Geim and Novoselov(2007)]{Geim2007}
A.~K. Geim and K.~S. Novoselov.
\newblock {The rise of graphene}.
\newblock \emph{Nature Materials}, 6\penalty0 (3):\penalty0 183--191, mar 2007.
\newblock ISSN 1476-1122.
\newblock \doi{10.1038/nmat1849}.

\bibitem[Xu et~al.(2016)Xu, Cheng, Du, Yang, Yu, Luo, Yin, Li, Dong, Ye,
  et~al.]{xu2016contacts}
Yang Xu, Cheng Cheng, Sichao Du, Jianyi Yang, Bin Yu, Jack Luo, Wenyan Yin,
  Erping Li, Shurong Dong, Peide Ye, et~al.
\newblock Contacts between two-and three-dimensional materials: ohmic,
  schottky, and p--n heterojunctions.
\newblock \emph{ACS nano}, 10\penalty0 (5):\penalty0 4895--4919, 2016.

\bibitem[Wang et~al.(2013)Wang, Cheng, Xu, Tsang, and Xu]{Wang2013a}
Xiaomu Wang, Zhenzhou Cheng, Ke~Xu, Hon~Ki Tsang, and Jian-Bin Xu.
\newblock {High-responsivity graphene/silicon-heterostructure waveguide
  photodetectors}.
\newblock \emph{Nature Photonics}, 7\penalty0 (11):\penalty0 888--891, nov
  2013.
\newblock ISSN 1749-4885.
\newblock \doi{10.1038/nphoton.2013.241}.

\bibitem[Li et~al.(2015)Li, Tian, Zhang, Singh, Du, Gu, Han, and Zhang]{Li2015}
Quan Li, Zhen Tian, Xueqian Zhang, Ranjan Singh, Liangliang Du, Jianqiang Gu,
  Jiaguang Han, and Weili Zhang.
\newblock {Active graphene-silicon hybrid diode for terahertz waves}.
\newblock \emph{Nature Communications}, 6\penalty0 (May), 2015.
\newblock ISSN 20411723.

\bibitem[He et~al.(2012)He, Lin, Liu, Zhu, Zhao, Chan, and Yan]{He2012}
Rong~Xiang He, Peng Lin, Zhi~Ke Liu, Hong~Wei Zhu, Xing~Zhong Zhao, Helen~L.W.
  Chan, and Feng Yan.
\newblock {Solution-gated graphene field effect transistors integrated in
  microfluidic systems and used for flow velocity detection}.
\newblock \emph{Nano Letters}, 12\penalty0 (3):\penalty0 1404--1409, 2012.
\newblock ISSN 15306984.
\newblock \doi{10.1021/nl2040805}.

\bibitem[Li et~al.(2010)Li, Zhu, Wang, Cao, Wei, Li, Jia, Li, Li, and
  Wu]{Li2010c}
Xinming Li, Hongwei Zhu, Kunlin Wang, Anyuan Cao, Jinquan Wei, Chunyan Li,
  Yi~Jia, Zhen Li, Xiao Li, and Dehai Wu.
\newblock {Graphene-on-silicon schottky junction solar cells}.
\newblock \emph{Advanced Materials}, 22\penalty0 (25):\penalty0 2743--2748,
  2010.
\newblock ISSN 09359648.
\newblock \doi{10.1002/adma.200904383}.

\bibitem[{Di Bartolomeo}(2016)]{DiBartolomeo2016}
Antonio {Di Bartolomeo}.
\newblock {Graphene Schottky diodes: An experimental review of the rectifying
  graphene/semiconductor heterojunction}.
\newblock \emph{Physics Reports}, 606:\penalty0 1--58, 2016.
\newblock ISSN 03701573.

\bibitem[Doukas et~al.(2022)Doukas, Mensz, Myoung, Ferrari, Goykhman, and
  Lidorikis]{doukas2022thermionic}
S~Doukas, P~Mensz, N~Myoung, AC~Ferrari, I~Goykhman, and E~Lidorikis.
\newblock Thermionic graphene/silicon schottky infrared photodetectors.
\newblock \emph{Physical Review B}, 105\penalty0 (11):\penalty0 115417, 2022.

\bibitem[Ono and Im(2022)]{ono2022thermal}
Yuzuki Ono and Hojun Im.
\newblock Thermal annealing effects on graphene/n-si schottky junction solar
  cell: Removal of pmma residues.
\newblock \emph{arXiv preprint arXiv:2209.08365}, 2022.

\bibitem[Wu et~al.(2022)Wu, Zhu, Tian, Yan, Liu, Xu, Xing, and
  Ren]{wu2022vertical}
Fan Wu, Zheng-Qiang Zhu, He~Tian, Zhaoyi Yan, Yanming Liu, Yang Xu, Chao-Yang
  Xing, and Tian-ling Ren.
\newblock Vertical wse2/bp/mos2 heterostructures with tunneling behaviors and
  photodetection.
\newblock \emph{Applied Physics Letters}, 121\penalty0 (11):\penalty0 113508,
  2022.

\bibitem[Geng et~al.(2022)Geng, Zhang, Ren, Dun, Li, Jin, Wang, Wu, Xie, Tian,
  et~al.]{geng2022directly}
Xiangshun Geng, Peigen Zhang, Jun Ren, Guan-Hua Dun, Yuanyuan Li, Jialun Jin,
  Chaolun Wang, Xing Wu, Dan Xie, He~Tian, et~al.
\newblock Directly integrated mixed-dimensional van der waals
  graphene/perovskite heterojunction for fast photodetection.
\newblock \emph{InfoMat}, 4\penalty0 (8):\penalty0 e12347, 2022.

\bibitem[Chen et~al.(2011)Chen, Aykol, Chang, Levi, and Cronin]{Chen2011}
Chun-Chung Chen, Mehmet Aykol, Chia-Chi Chang, A.~F.~J. Levi, and Stephen~B.
  Cronin.
\newblock {Graphene-Silicon Schottky Diodes}.
\newblock \emph{Nano Letters}, 11\penalty0 (11):\penalty0 5097--5097, 2011.
\newblock ISSN 1530-6984.

\bibitem[An et~al.(2013)An, Liu, and Kar]{An2013}
Xiaohong An, Fangze Liu, and Swastik Kar.
\newblock {Optimizing performance parameters of graphene-silicon and thin
  transparent graphite-silicon heterojunction solar cells}.
\newblock \emph{Carbon}, 57:\penalty0 329--337, 2013.
\newblock ISSN 00086223.

\bibitem[Zhang et~al.(2013)Zhang, Xie, Jie, Zhang, Wu, and
  Zhang]{zhang2013high}
Xiaozhen Zhang, Chao Xie, Jiansheng Jie, Xiwei Zhang, Yiming Wu, and Wenjun
  Zhang.
\newblock High-efficiency graphene/si nanoarray schottky junction solar cells
  via surface modification and graphene doping.
\newblock \emph{Journal of Materials Chemistry A}, 1\penalty0 (22):\penalty0
  6593--6601, 2013.

\bibitem[Wu et~al.(2013)Wu, Zhang, Jie, Xie, Zhang, Sun, Wang, and
  Gao]{wu2013graphene}
Yiming Wu, Xiaozhen Zhang, Jiansheng Jie, Chao Xie, Xiwei Zhang, Baoquan Sun,
  Yan Wang, and Peng Gao.
\newblock Graphene transparent conductive electrodes for highly efficient
  silicon nanostructures-based hybrid heterojunction solar cells.
\newblock \emph{The Journal of Physical Chemistry C}, 117\penalty0
  (23):\penalty0 11968--11976, 2013.

\bibitem[Shi et~al.(2013)Shi, Li, Yang, Zhang, Li, Li, Shang, Wu, Li, Wei,
  et~al.]{shi2013colloidal}
Enzheng Shi, Hongbian Li, Long Yang, Luhui Zhang, Zhen Li, Peixu Li, Yuanyuan
  Shang, Shiting Wu, Xinming Li, Jinquan Wei, et~al.
\newblock Colloidal antireflection coating improves graphene--silicon solar
  cells.
\newblock \emph{Nano letters}, 13\penalty0 (4):\penalty0 1776--1781, 2013.

\bibitem[Wang et~al.(2019)Wang, Gu, Zhang, and Chen]{Wang2019h}
Bo~Wang, Yijia Gu, Shujun Zhang, and Long~Qing Chen.
\newblock {Flexoelectricity in solids: Progress, challenges, and perspectives}.
\newblock \emph{Progress in Materials Science}, 106\penalty0 (May), 2019.
\newblock ISSN 00796425.

\bibitem[Kogan(1964)]{kogan1964piezoelectric}
Sh~M Kogan.
\newblock Piezoelectric effect during inhomogeneous deformation and acoustic
  scattering of carriers in crystals.
\newblock \emph{Soviet Physics-Solid State}, 5\penalty0 (10):\penalty0
  2069--2070, 1964.

\bibitem[Tagantsev(1986)]{Tagantsev1986}
A.~K. Tagantsev.
\newblock {Piezoelectricity and flexoelectricity in crystalline dielectrics}.
\newblock \emph{Physical Review B}, 34\penalty0 (8):\penalty0 5883--5889, 1986.
\newblock ISSN 01631829.

\bibitem[Wang et~al.(2020)Wang, Liu, Feng, Zhang, Zhu, Zhai, Qin, and
  Wang]{Wang2020o}
Longfei Wang, Shuhai Liu, Xiaolong Feng, Chunli Zhang, Laipan Zhu, Junyi Zhai,
  Yong Qin, and Zhong~Lin Wang.
\newblock {Flexoelectronics of centrosymmetric semiconductors}.
\newblock \emph{Nature Nanotechnology}, 15\penalty0 (8):\penalty0 661--667,
  2020.
\newblock ISSN 17483395.

\bibitem[Park et~al.(2021{\natexlab{a}})Park, Wang, Chen, Noh, Yang, and
  Lee]{Park2021a}
Sung~Min Park, Bo~Wang, Long~Qing Chen, Tae~Won Noh, Sang~Mo Yang, and Daesu
  Lee.
\newblock {Flexoelectric control of physical properties by atomic force
  microscopy}.
\newblock \emph{Applied Physics Reviews}, 8\penalty0 (4), 2021{\natexlab{a}}.
\newblock ISSN 19319401.

\bibitem[Naumov et~al.(2009)Naumov, Bratkovsky, and Ranjan]{Naumov2009}
Ivan Naumov, Alexander~M. Bratkovsky, and V.~Ranjan.
\newblock {Unusual flexoelectric effect in two-dimensional noncentrosymmetric
  sp2-bonded crystals}.
\newblock \emph{Physical Review Letters}, 102\penalty0 (21):\penalty0 2--5,
  2009.
\newblock ISSN 00319007.
\newblock \doi{10.1103/PhysRevLett.102.217601}.

\bibitem[Springolo et~al.(2021)Springolo, Royo, and Stengel]{Springolo2020}
Matteo Springolo, Miquel Royo, and Massimiliano Stengel.
\newblock {Direct and Converse Flexoelectricity in Two-Dimensional Materials}.
\newblock \emph{Physical Review Letters}, 127\penalty0 (21):\penalty0 216801,
  nov 2021.
\newblock ISSN 0031-9007.
\newblock \doi{10.1103/PhysRevLett.127.216801}.

\bibitem[Park et~al.(2021{\natexlab{b}})Park, Wang, Chen, Noh, Yang, and
  Lee]{Park2021}
Sung~Min Park, Bo~Wang, Long-Qing Chen, Tae~Won Noh, Sang~Mo Yang, and Daesu
  Lee.
\newblock {Flexoelectric control of physical properties by atomic force
  microscopy}.
\newblock \emph{Applied Physics Reviews}, 8\penalty0 (4):\penalty0 041327,
  2021{\natexlab{b}}.

\bibitem[Jiang et~al.(2022)Jiang, Wang, Wang, Zhang, Niu, Deng, Xu, Lun, Liu,
  Xia, Lu, and Hong]{Jiang2022}
Xingan Jiang, Xueyun Wang, Xiaolei Wang, Xiangping Zhang, Ruirui Niu, Jianming
  Deng, Sheng Xu, Yingzhuo Lun, Yanyu Liu, Tianlong Xia, Jianming Lu, and
  Jiawang Hong.
\newblock {Manipulation of current rectification in van der Waals ferroionic
  CuInP2S6}.
\newblock \emph{Nature Communications}, 13\penalty0 (1):\penalty0 574, dec
  2022.
\newblock ISSN 2041-1723.

\bibitem[Wang et~al.(2021)Wang, Zhou, Cui, Deng, Xu, Xu, Ye, Jiang, Shang, Zhu,
  Zhang, Li, Hu, and Chu]{Wang2021e}
Xiang Wang, Xin Zhou, Anyang Cui, Menghan Deng, Xionghu Xu, Liping Xu, Yan Ye,
  Kai Jiang, Liyan Shang, Liangqing Zhu, Jinzhong Zhang, Yawei Li, Zhigao Hu,
  and Junhao Chu.
\newblock {Flexo-photoelectronic effect in n-type/p-type two-dimensional
  semiconductors and a deriving light-stimulated artificial synapse}.
\newblock \emph{Materials Horizons}, 8\penalty0 (7):\penalty0 1985--1997, 2021.
\newblock ISSN 20516355.

\bibitem[Wang et~al.(2015)Wang, Tian, Xie, Shu, Mi, Ali~Mohammad, Xie, Yang,
  Xu, and Ren]{wang2015observation}
Xiaomu Wang, He~Tian, Weiguang Xie, Yi~Shu, Wen-Tian Mi, Mohammad Ali~Mohammad,
  Qian-Yi Xie, Yi~Yang, Jian-Bin Xu, and Tian-Ling Ren.
\newblock Observation of a giant two-dimensional band-piezoelectric effect on
  biaxial-strained graphene.
\newblock \emph{NPG Asia Materials}, 7\penalty0 (1):\penalty0 e154--e154, 2015.

\bibitem[Sun et~al.(2021)Sun, Zhu, Zhang, Chen, and Wang]{Sun2021}
Liang Sun, Lifeng Zhu, Chunli Zhang, Weiqiu Chen, and Zhonglin Wang.
\newblock {Mechanical Manipulation of Silicon-based Schottky Diodes via
  Flexoelectricity}.
\newblock \emph{Nano Energy}, 83\penalty0 (December 2020):\penalty0 105855,
  2021.
\newblock ISSN 22112855.

\bibitem[Yang et~al.(2018)Yang, Kim, and Alexe]{Yang2018}
Ming-min Yang, Dong~Jik Kim, and Marin Alexe.
\newblock {Flexo-photovoltaic effect}.
\newblock \emph{Science}, 360\penalty0 (6391):\penalty0 904--907, may 2018.
\newblock ISSN 0036-8075.

\bibitem[Shu et~al.(2020)Shu, Ke, Fei, Huang, Wang, Gong, Jiang, Wang, Li, Lei,
  Rao, Zhou, Zheng, Yao, Wang, Stengel, and Catalan]{Shu2020}
Longlong Shu, Shanming Ke, Linfeng Fei, Wenbin Huang, Zhiguo Wang, Jinhui Gong,
  Xiaoning Jiang, Li~Wang, Fei Li, Shuijin Lei, Zhenggang Rao, Yangbo Zhou,
  Ren~Kui Zheng, Xi~Yao, Yu~Wang, Massimiliano Stengel, and Gustau Catalan.
\newblock {Photoflexoelectric effect in halide perovskites}.
\newblock \emph{Nature Materials}, 19\penalty0 (6):\penalty0 605--609, 2020.
\newblock ISSN 14764660.

\bibitem[Zhang et~al.(2019)Zhang, Ideue, Onga, Qin, Suzuki, Zak, Tenne, Smet,
  and Iwasa]{Zhang2019d}
Y.~J. Zhang, T.~Ideue, M.~Onga, F.~Qin, R.~Suzuki, A.~Zak, R.~Tenne, J.~H.
  Smet, and Y.~Iwasa.
\newblock {Enhanced intrinsic photovoltaic effect in tungsten disulfide
  nanotubes}.
\newblock \emph{Nature}, 570\penalty0 (7761):\penalty0 349--353, 2019.
\newblock ISSN 14764687.

\bibitem[Jiang et~al.(2021)Jiang, Chen, Hu, Xiang, Zhang, Wang, Wang, and
  Shi]{Jiang2021}
Jie Jiang, Zhizhong Chen, Yang Hu, Yu~Xiang, Lifu Zhang, Yiping Wang, Gwo-ching
  Wang, and Jian Shi.
\newblock {Flexo-photovoltaic effect in MoS2}.
\newblock \emph{Nature Nanotechnology}, 16\penalty0 (8):\penalty0 894--901, aug
  2021.
\newblock ISSN 1748-3387.
\newblock \doi{10.1038/s41565-021-00919-y}.

\bibitem[Peng et~al.(2022)Peng, Liu, Du, Bodepudi, Li, Liu, Lai, Cao, Fang,
  Liu, Liu, Lv, Abid, Liu, Jin, Wu, Lin, Cong, Tan, Zhu, Xiong, Wang, Hu, Duan,
  Yu, Xu, Xu, and Gao]{Peng2022a}
Li~Peng, Lixiang Liu, Sichao Du, Srikrishna~Chanakya Bodepudi, Lingfei Li, Wei
  Liu, Runchen Lai, Xiaoxue Cao, Wenzhang Fang, Yingjun Liu, Xinyu Liu,
  Jianhang Lv, Muhammad Abid, Junxue Liu, Shengye Jin, Kaifeng Wu, Miao‐Ling
  Lin, Xin Cong, Ping‐Heng Tan, Haiming Zhu, Qihua Xiong, Xiaomu Wang, Weida
  Hu, Xiangfeng Duan, Bin Yu, Zhen Xu, Yang Xu, and Chao Gao.
\newblock {Macroscopic assembled graphene nanofilms based room temperature
  ultrafast mid‐infrared photodetectors}.
\newblock \emph{InfoMat}, 4\penalty0 (6):\penalty0 1--12, jun 2022.
\newblock ISSN 2567-3165.
\newblock \doi{10.1002/inf2.12309}.
\newblock URL \url{https://onlinelibrary.wiley.com/doi/10.1002/inf2.12309}.

\bibitem[Liang et~al.(2017)Liang, Hu, {Di Bartolomeo}, Adam, and
  Ang]{Liang2017}
Shi~Jun Liang, Wei Hu, A.~{Di Bartolomeo}, Shaffique Adam, and Lay~Kee Ang.
\newblock {A modified Schottky model for graphene-semiconductor (3D/2D)
  contact: A combined theoretical and experimental study}.
\newblock \emph{Technical Digest - International Electron Devices Meeting,
  IEDM}, pages 14.4.1--14.4.4, 2017.
\newblock ISSN 01631918.

\end{thebibliography}
\bibliographystyle{unsrtnat}
\end{document}